\documentclass[10pt,conference]{IEEEtran}
\IEEEoverridecommandlockouts

\usepackage{comment}
\usepackage{cite}
\usepackage{amsmath,amssymb,amsfonts}
\usepackage{algorithmic}
\usepackage{graphicx}
\usepackage{textcomp}
\usepackage{xcolor}
\usepackage{xspace}
\usepackage{listings}
\usepackage{url}
\usepackage{todonotes}
\usepackage{multirow}
\usepackage{paralist}
\usepackage{svg}

\def\BibTeX{{\rm B\kern-.05em{\sc i\kern-.025em b}\kern-.08em
    T\kern-.1667em\lower.7ex\hbox{E}\kern-.125emX}}

\newboolean{showcomments}
\setboolean{showcomments}{true}         
\ifthenelse{\boolean{showcomments}}
{\newcommand{\nb}[2]{
		\fbox{\bfseries\sffamily\scriptsize#1}
		{\sf\small$\blacktriangleright$\textit{\textcolor{red}{#2}}$\blacktriangleleft$}
	}
}
{\newcommand{\nb}[2]{}
	
}

\newcommand{\jdoctor}{\textsc{JDoc\-tor}\xspace}
\newcommand{\evosuite}{\textsc{Evo\-suite}\xspace}
\newcommand{\jbmc}{\textsc{JBMC}\xspace}
\newcommand{\pex}{\textsc{Pex}\xspace}
\newcommand{\technique}{\textsc{JuDoT}\xspace}
\newcommand{\baseline}{\textsc{Evo\-suite+JDoc\-tor}\xspace}
\newcommand{\code}[1]{\text{\texttt{#1}}}

\begin{document}

\title{Automated Test Generation from Program Documentation Encoded in
Code Comments\thanks{This work is partially supported by the PRIN 2022 project Big Sistah,  2022EYX28N, and by the PNRR project SOP, H73C22000890001 (part of SERICS PE00000014).

\copyright2025 IEEE.  Personal use of this material is permitted.  Permission from IEEE must be obtained for all other uses, in any current or future media, including reprinting/republishing this material for advertising or promotional purposes, creating new collective works, for resale or redistribution to servers or lists, or reuse of any copyrighted component of this work in other works.}}

\author{\IEEEauthorblockN{Giovanni Denaro, Luca Gugliemo}
\IEEEauthorblockA{\textit{Dept. of Informatics, Systems and Communication, University of Milano-Bicocca}, Milano, Italy\\
\{giovanni.denaro, luca.guglielmo\}@unimib.it}
}

\maketitle

\begin{abstract}
Documenting the functionality of  software units with code comments, e.g., Javadoc comments, is a common programmer best-practice in software engineering.
This paper introduces a novel test generation technique that exploits the code-comment documentation  constructively.
We originally address those behaviors as test objectives, which we pursue in search-based fashion. We 
deliver test cases 
with names and oracles properly contextualized on the target behaviors.
Our experiments against a benchmark of 118 Java classes indicate that the proposed approach successfully tests many software behaviors that may remain untested with coverage-driven test generation approaches, and distinctively detects unknown failures.

\end{abstract}

\begin{IEEEkeywords}
Automated test generation, Search based testing, Testing code documentation, Test oracles.
\end{IEEEkeywords}


\section{Introduction}

Documenting the intended functionality of software units with code comments 
is a common programmer best-practice in software engineering~\cite{savitch2016absolute,sommerville2011software}. 
Over time this practice evolved into standards and tooling support for writing  code documentation with semi-structured comments 
Notable examples are the Javadoc and PerlPod 
markup languages, for documenting programs in Java and Perl, respectively~\cite{javadoc,perlpod}. 
The Doxygen documentation generator generalizes the approach across several programming languages~\cite{noauthor_doxygen_nodate}.
Many Java IDEs automatically insert templates for Javadoc comments~\cite{eclipseide}.

Recently, the \jdoctor~\cite{Blasi:Jdoctor:ISSTA:2018} approach proposed that Javadoc comments can be turned to executable specifications (precondition and postcondition contracts), 
and then used as test oracles, in order to address the oracle problem in test case generation~\cite{barr2014oracle}. 
\jdoctor's authors proposed to use their technique 
to strengthen the current mainstream test generation techniques, 
which synthesize unit-level test cases by incrementally satisfying 
test objectives that arise from the structure of the code,
but miss the ability of equipping the test cases with proper oracles. 
This is the case for example of test generation techniques based on symbolic execution, random testing or search-based testing~\cite{godefroid:dart:pldi:2005,anand_jpf-se_2007,Pacheco:Randoop:ICSE:2007,fraser_whole_2013,Saswat:JPF-SE:TACAS:2007,braione_symbolic_2015,braione_jbse_2016,Braione:SUSHI:ISSTA:2017,mcminn_search-based_2004,tonella_evolutionary_2004,baldoni:symsurvey:acm:2018,harman:searchtrends:acmsurvey:2012}.
Without proper oracles, those techniques result  most often in test suites that  suffer from  false alarms (fail against correct programs) or missed alarms (pass against faulty programs)
\jdoctor mitigates those issues.

This paper observes that, regardless of any improvement on the oracles,
  code-coverage-driven test generators end up by-design with testing a selected set of software behaviors: the ones that  coincidentally relate with increasing code coverage.   
 However, 
many relevant software behaviors may remain untested, despite being documented by the developers.
We further discuss this limitation with examples in Section 2. In Section 4, while using the test generator \evosuite against a benchmark of 118 Java classes, we provide empirical evidence that this phenomenon has significant impacts.

Motivated by the above observation, this paper 
introduces a novel test generation technique that exploits the code-comment documentation constructively.
We tackle the test generation problem in the style of functional test design~\cite{pezze2008software,Ostrand:CategoryPartition:ACM:1988}:
we first identify the relevant program behaviors encompassed in the documentation, and  then 
engage in generating test cases that address those behaviors effectively. 

We instantiated our technique for programs written in Java and documented with Javadoc, by implementing the prototype \technique (\underline{Ju}nit-from-Java\underline{Do}c \underline{T}est generation).
\technique grounds on the available functionality of \jdoctor 
for inferring 
precondition and postcondition contracts  from the Javadoc comments. But, differently than \jdoctor, \technique originally leverages those contracts  as  first-class test objectives, which it pursues in search-based fashion in order to generate test cases. 
For each contract, \technique aims to exercise the program behaviors specified in the contract,
but also considers the goal of detecting possible failures with respect to the contract.
\technique 
also tracks each generated test case to 
the corresponding \emph{focal contract}, and yields test cases that are specifically contextualized: they
include the respective focal contracts as test oracles, and have  names mimicking the natural-language wording of the focal contracts. 
This addresses another common pitfall of code-driven test generators, which generally produce test cases with generic names (e.g., \texttt{test0}).  

We experimented with \technique against a benchmark of 118 Java classes. The results indicate that our approach could successfully test many more contracts  than a traditional search-based test generator, \evosuite, that we considered as baseline. We measured an increase of 14.8\% 
in the number of tested contracts with respect to the baseline, which we regard as a significant improvement.
Moreover, the test cases from our technique revealed 107 contract violations, 31 additional violations compared to  the baseline.
By manual inspection, we further tracked those violations to 45 actual failures of the classes under test (the other ones were false alarms due to imprecise inferences while processing the natural language), confirming the effectiveness of the approach.

The rest of this paper is organized as follows. Section~\ref{sec:overview} exemplifies the limitations of current test generators on a working example, and provides an overview of the characteristics of our approach. Section~\ref{sec:approach} presents \technique in detail. Section~\ref{sec:experiments} reports our experimental evaluation of \technique against a benchmark of 118 Java classes. Section~\ref{sec:related} surveys the related work. Section~\ref{sec:conclusions} summarizes our conclusions and outlines our plans for further research on this topic.


\section{Motivating Examples}
\label{sec:overview}

In this section we exemplify code-comment documentation in Javadoc for
a sample Java program, further discuss
the limits of current test generators to
thoroughly exercise the behaviors in the sample program, and hint to the original contributions of the test generation technique proposed in this paper. 

\subsection{Working Example}
Figure~\ref{fig:sampleprog} excerpts a sample Java class, class \code{GiftPack}, that includes code-comment documentation in Javadoc.
The comments at lines \ref{fig:sampleprog:tag:param}--\ref{fig:sampleprog:tag:return} specify the expected behavior of method \code{unwrapAndSave}, by indicating contracts on preconditions and postconditions.
The method code starts 
at line~\ref{fig:sampleprog:method:start}. 

In a nutshell, the method should take the content of the \code{GiftPack} (i.e., the attribute \code{gift}, line~\ref{fig:sampleprog:gift}) and add it (line~\ref{fig:sampleprog:method:addgift}) in the \code{drawer} object that it receives as parameter (line~\ref{fig:sampleprog:method:start}). 
The Javadoc comment at line~\ref{fig:sampleprog:tag:param} references the input parameter \code{limit} with the tag \code{@param},  and
specifies the precondition  contract  
that parameter \code{limit} must have a strictly positive value. 
The Javadoc comments at lines~\ref{fig:sampleprog:tag:throws:gift}, \ref{fig:sampleprog:tag:throws:drawer} and \ref{fig:sampleprog:tag:throws:nomore} specify exceptional-behavior postconditions (tag \code{@throws}) for the cases in which the attribute \code{gift} is null, the parameter \code{drawer} is null, or the \code{drawer} exceeds the given \code{limit}; accordingly the method shall throw  \code{EmptyException}, \code{NullPointerException} and \code{NoMoreException}, respectively. The Javadoc comment at line~\ref{fig:sampleprog:tag:return} specifies a postcondition on the return value of the method (tag \code{@return}): it shall return $false$ if the
\code{gift} object is already contained in the \code{drawer}, or $true$ otherwise.

\begin{figure}[t!]
    \centering
\begin{lstlisting}[language=java,numbers=left,tabsize=2,
    stepnumber=1, basicstyle=\scriptsize,xleftmargin=3.5em,escapechar=|,breaklines=true, postbreak=\mbox{$\hookrightarrow$}\space]
class GiftPack {
	private Something gift;|\label{fig:sampleprog:gift}|

	GiftPack(Something st) {this.gift = st;}
	Something getGift() {return gift;}

	/**
	 * @param limit an int, must be positive |\label{fig:sampleprog:tag:param}|
	 * @throws EmptyException if gift is null |\label{fig:sampleprog:tag:throws:gift}|
	 * @throws NullPointerException if drawer is null |\label{fig:sampleprog:tag:throws:drawer}|
	 * @throws NoMoreException if the drawer |~~~~~~~~|exceeds the limit |\label{fig:sampleprog:tag:throws:nomore}|
	 * @return false if the drawer already contains the gift, true otherwise |\label{fig:sampleprog:tag:return}|
	 */
	boolean unwrapAndSave(Drawer drawer, int limit) { |\label{fig:sampleprog:method:start}|
		checkValidLimit(limit); |\label{fig:sampleprog:method:log}|
		if (drawer.exceeds(limit)) |\label{fig:sampleprog:method:nullpointer}|
			throw new NoMoreException();|\label{fig:sampleprog:method:nomore}|
		drawer.add(gift);  |\label{fig:sampleprog:method:addgift}|
		return true; |\label{fig:sampleprog:method:return}|
	}
	void checkValidLimit(int limit) {
		if (limit <= 0) logWarning("bad limit"); 
	}
}
\end{lstlisting} 
    \caption{A sample Java class with Javadoc documentation}
    \label{fig:sampleprog}
\end{figure}

Being knowledgeable of the above contracts, we can  observe that the current implementation of method \code{unwrapAndSave} in the figure is flawed. 
The current implementation does not guarantee neither the 
\code{@throws}-contract at line~\ref{fig:sampleprog:tag:throws:gift}, as 
line~\ref{fig:sampleprog:method:addgift} can receive attribute \code{gift} with null value,  
nor the \code{@return}-contract at line~\ref{fig:sampleprog:tag:return},  
as line~\ref{fig:sampleprog:method:return} returns $true$ regardless of any content of the \code{drawer}. 
On the other hand, the implementation complies with the contracts at line~\ref{fig:sampleprog:tag:throws:drawer} and~\ref{fig:sampleprog:tag:throws:nomore}, as the execution of line~\ref{fig:sampleprog:method:nullpointer} indeed throws a null-pointer exception if \code{drawer} is null, and line~\ref{fig:sampleprog:method:nomore} throws an exception of type \code{NoMoreException} upon detecting that \code{drawer} \emph{exceeds} \code{limit}. 
We also observe that 
the implementation handles the precondition on parameter \code{limit}
 by logging a warning message (line~\ref{fig:sampleprog:method:log}). This is a valid (though possibly not robust) implementation:  guaranteeing the preconditions is responsibility of callers and, 
 if a precondition is violated, the behavior of the implementation can be any. 

\subsection{Limits of Current Test Generation Approaches}

At the state of the art, the mainstream
  test generation techniques
   synthesize unit-level test cases by addressing 
test objectives that arise from the structure of the code, e.g., aiming to execute the statements, the branches, the paths, the data flow relations, the mutations in the code, or selections and combinations of those types of test objectives. Those techniques generally build on the classic approaches of random testing, search-based testing or symbolic execution, or possibly variations or combinations of those main approaches~\cite{godefroid:dart:pldi:2005,anand_jpf-se_2007,Pacheco:Randoop:ICSE:2007,fraser_whole_2013,Saswat:JPF-SE:TACAS:2007,braione_symbolic_2015,braione_jbse_2016,Braione:SUSHI:ISSTA:2017,mcminn_search-based_2004,tonella_evolutionary_2004,baldoni:symsurvey:acm:2018,harman:searchtrends:acmsurvey:2012}.

However, as a result of their code-coverage-driven nature, those test generators may often fail to thoroughly 
assess 
the programs with respect to the documented 
expectations, incurring \emph{untested behaviors}, or 
 the \emph{oracle problem}, or both.

For instance, below we exemplify  the case of untested behaviors and the oracle problem with reference to the test cases that the test generator \evosuite~\cite{fraser_whole_2013} produces for the class \code{GiftPack} of Figure~\ref{fig:sampleprog}. \evosuite synthesizes test cases in search-based fashion driven by fitness functions that can be set to address several code-coverage criteria. 
For the class \code{GiftPack}, any run of \evosuite quickly converges in generating test suites that execute all program statements, all branches, all exceptions and all mutants\footnote{\evosuite synthesizes mutants by replacing variables, constants, comparison operators, bitwise operators and arithmetic operators in the program.} in the code.
 Figure~\ref{fig:evotests} shows a test suite generated with \evosuite.

\begin{figure}[t!]
    \centering
\begin{lstlisting}[language=java,numbers=left,tabsize=2,
    stepnumber=1, basicstyle=\scriptsize,xleftmargin=3.5em,escapechar=|,breaklines=true, postbreak=\mbox{$\hookrightarrow$}\space]
void test0() {
    Something s0 = new Something();
    GiftPack gp0 = new GiftPack(s0);
    try { 
        giftPack0.unwrapAndSave(null, -1000);|\label{fig:evotests:negative}|
        fail("Expecting NullPointerException");|\label{fig:evotests:npe}|
    } catch(NullPointerException e) {}
}
void test1() {
    Something s0 = new Something();
    GiftPack gp0 = new GiftPack(s0);
    Drawer d0 = new Drawer();
    try { 
        giftPack0.unwrapAndSave(d0, 0);|\label{fig:evotests:zero}|
        fail("Expecting RuntimeException");|\label{fig:evotests:rte}|
    } catch(RuntimeException e) {}
}
void test2() {
      GiftPack gp0 = new GiftPack(null);|\label{fig:evotests:null}|
      Drawer d0 = new Drawer();
      boolean boolean0 = 
        giftPack0.unwrapAndSave(d0, 3184);
      assertTrue(boolean0);|\label{fig:evotests:true}|
}
\end{lstlisting} 
    \caption{Test cases computed with \evosuite for class GiftPack}
    \label{fig:evotests}
\end{figure}

\paragraph{Untested Behaviors}

Code-coverage-driven test generators incrementally select test cases that execute additional elements of the target coverage domain(s). 
However, as the relation between the code elements and the possible program behaviors  is coincidental,
they may easily miss some behaviors that do not distinctively map to a target code element. 

For instance, the class \code{GiftPack} in Figure~\ref{fig:sampleprog} has a single line of code (line~\ref{fig:sampleprog:method:addgift}) at which the program adds the current \code{gift} to the \code{drawer}. 
As a result, a code-coverage-driven test generator like \evosuite  might end up with providing a single test case that executes line~\ref{fig:sampleprog:method:addgift}, e.g., the test case \code{test2} in Figure~\ref{fig:evotests}.
\evosuite has no explicit guidance for generating a test case in which \code{gift} initially belongs to \code{drawer}. It may in principle generate a test case like this, but only by pure chance.
In our experience, out of 30 repetitions, \evosuite consistently missed to generate any test case in which \code{gift} belongs to \code{drawer}. Thus, it could not spot the flaw of the sample implementation that returns $true$ also in that case (in contrast with the Javadoc contract documented at line~\ref{fig:sampleprog:tag:return}).

Other reasons for which a test generator may incur untested behaviors, encompassing technical limitations on synthesizing some types of inputs, are out of the scope of 
this paper. We focus on untested behaviors that derive from lack of guidance.

\paragraph{The Oracle Problem}
The oracle problem concerns the hardness of setting the generated test cases with proper, automated test oracles~\cite{barr2014oracle}. For instance, the test oracles could be provided in the form of assertion statements that
check if the results at runtime are consistent with the expectations known for the program under test, and make the test cases produce test reports (i.e., pass or fail) accordingly. 

Unfortunately, code-coverage-driven test generators generally know nothing about the actual expectations, and they can only fall down to synthesizing generic, specification-agnostic oracles~\cite{Fraser:EvoSuite:ESECFSE:2011,Pacheco:Randoop:ICSE:2007,kurian2023automatically}. For example, a common approach is to produce oracles that simply accept the current results of all test cases as the correct ones, as  oracles of this type can become useful lately for regression testing purposes.
In fact, this is the approach of \evosuite, that readers can observe in Figure~\ref{fig:evotests}: \code{test0} and \code{test1} are set for checking (at line~\ref{fig:evotests:npe} and line~\ref{fig:evotests:rte}, respectively) that the program shall throw the exceptions that the current implementation is throwing;
  \code{test2} checks (at line~\ref{fig:evotests:true}) that the program shall return $true$, as \evosuite observed that the current implementation returned $true$. 

Due to ignoring the actual expectations, test generators often incur the synthesis of
test cases that either are invalid, because they violate the preconditions of the program, or 
result in either false alarms or missed alarms~\cite{Blasi:Jdoctor:ISSTA:2018}.
 For example,  the test cases \code{test0} and \code{test1} in Figure~\ref{fig:evotests} are invalid, as they execute method \code{unwrapAndSave} with parameter \code{limit} set to a negative value (line~\ref{fig:evotests:negative}) and zero (line~\ref{fig:evotests:zero}), respectively, which contrasts with the precondition of the method (Figure~\ref{fig:sampleprog}, line~\ref{fig:sampleprog:tag:param}). 
The test case \code{test2} passes, but in fact it incurs a missed alarm, as the specification (Figure~\ref{fig:sampleprog}, line~\ref{fig:sampleprog:tag:throws:gift}) requires that method \code{unwrapAndSave} should throw an \code{EmptyException} if the \code{gift} field  is set to null as in this case (Figure~\ref{fig:evotests}, line~\ref{fig:evotests:null}). 

\subsection{The Test Generator \technique by Example}

Figure~\ref{fig:jdttests} exemplifies a test suite that \technique, the  test generator  proposed in this paper, generated for the class \code{GiftPack} of Figure~\ref{fig:sampleprog}. The test cases focus on the Javadoc contracts, with test names that mimic the 
contract descriptions  and test oracles based on the expectations of the contracts. The test cases at line~\ref{fig:jdttests:fail:empty} and line~\ref{fig:jdttests:fail:contains} reveal the two bugs of the current implementation that we commented above. In particular the test case at line~\ref{fig:jdttests:fail:contains}, with name \emph{test\-Unwrap\-And\-Save\_False\-If\-Drawer\-Already\-Contains\-Gift\-True\-Otherwise\_0}, corresponds to the behavior that remained untested with the test suites from \evosuite: it calls method \code{unwrapAndSave} with a \code{gift} that already belongs to the current \code{drawer}.
This makes the current implementation fail as expected, revealing the bug.

\begin{figure}[t!]
    \centering
\begin{lstlisting}[language=java,numbers=left,tabsize=2,
    stepnumber=1, basicstyle=\scriptsize,xleftmargin=3.5em,escapechar=|,breaklines=true, postbreak=\mbox{$\hookrightarrow$}\space]
void testUnwrapAndSave_EmptyExIfGiftIsNull() {|\label{fig:jdttests:fail:empty}|
  GiftPack gp0 = new GiftPack(null);
  Drawer d0 = new Drawer();
  try {
    boolean b0 = gp0.unwrapAndSave(d0, 153);
    fail("Expected EmptyException");
  } catch (EmptyException e) { /* ok */ }
}
void testUnwrapAndSave_NullPointerExIfDrawerIsNull(){
  Something s0 = new Something();
  GiftPack gp0 = new GiftPack(s0);
  try {
    boolean b0 = gp0.unwrapAndSave(null, 207);
    fail("Expected NullPointerException");
  } catch (NullPointerException e) { /* ok */ }
}
void testUnwrapAndSave_NoMoreExIfDrawerExceedsLimit(){
  Something s0 = new Something();
  GiftPack gp0 = new GiftPack(s0);
  Drawer d0 = new Drawer();
  boolean b0 = gp0.unwrapAndSave(d0, 1);
  try {
    boolean b1 = gp0.unwrapAndSave(d0, 1);
    fail("Expected NoMoreException");
  } catch (NoMoreException e) { /* ok */ }
}
void testUnwrapAndSave_FalseIfDrawerAlreadyContains GiftTrueOtherwise_0(){|\label{fig:jdttests:fail:contains}|
  Something s0 = new Something();
  GiftPack gp0 = new GiftPack(s0);
  Drawer d0 = new Drawer();
  boolean b0 = d0.add(s0);
  boolean b1 = gp0.unwrapAndSave(d0, 153);
  assertFalse(b1);
}
void testUnwrapAndSave_FalseIfDrawerAlreadyContains GiftTrueOtherwise_1(){
  Something s0 = new Something();
  GiftPack gp0 = new GiftPack(s0);
  Drawer d0 = new Drawer();
  boolean b0 = gp0.unwrapAndSave(d0, 2);
  assertTrue(b0);
}
\end{lstlisting} 
    \caption{Test cases computed with \technique for class GiftPack}
    \label{fig:jdttests}
\end{figure}

\technique achieves this result by
exploiting the Javadoc specifications constructively in the test generation process. 
In \technique, the postcondition contracts inferred from the Javadoc comments are first-class test objectives that \technique pursues in search-based fashion. Thus, \technique tackles the untested-behavior issues discussed above.

By design, each test case from \technique focuses on a specific Javadoc contract 
that we refer to as \emph{the focal contract} of the test case.
\technique synthesizes test cases contextualized on the corresponding focal contracts, namely, with both the test names and
test oracles based on the focal contracts.
This can be observed in all test cases of Figure~\ref{fig:jdttests}.
Thus \technique contributes to address the oracle problem, and improves the readability of the test suites with respect to 
using generic names (e.g., \code{test0}, \code{test1} and \code{test2}, as in Figure~\ref{fig:evotests}).

\section{The \technique Technique} \label{sec:approach}
\begin{figure*}[h]
    \centering
    \includegraphics[width=\textwidth]{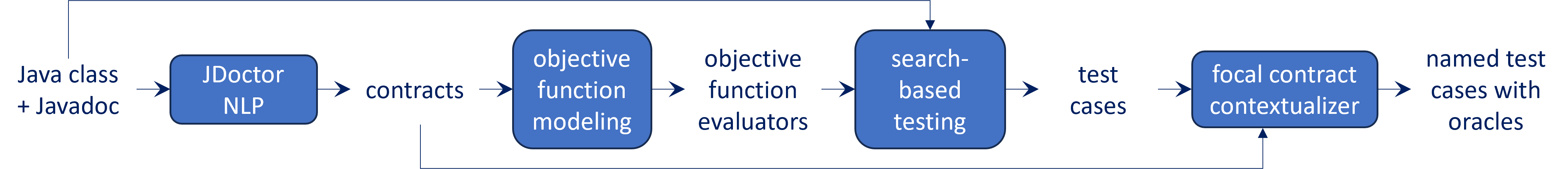}
    \caption{Workflow of \technique}
    \label{fig:workflow}
\end{figure*}

We sketch the main workflow of \technique in Figure~\ref{fig:workflow}. Given a Java class as input, \technique first executes \jdoctor to infer the contracts documented in the Javadoc comments (Section \ref{sec:processing:javadoc}). Then, it models each contract as an objective function and synthesizes evaluator programs for those objective functions (Section \ref{sec:objective:function}). Next,  it addresses the fitness functions with search based testing to generate test cases (Section \ref{sec:generating:test:cases}), and post-processes the test cases to add names and oracles based on the focal contracts (Section \ref{sec:contextualizing:test:cases:contracts}). Below, we explain each step in detail.

\subsection{Inferring Contracts from Javadoc (via \jdoctor)}
\label{sec:processing:javadoc}
As first step, \technique relies on \jdoctor~\cite{Blasi:Jdoctor:ISSTA:2018} to analyze the Javadoc contracts documented in the class under test.
\jdoctor renders those contracts in the form of \emph{precondition} and \emph{postcondition contracts}:
The preconditions correspond to 
 Javadoc comments marked with tag \texttt{@param}. The postconditions correspond to 
 Javadoc comments marked with either tag \texttt{@throws} or tag \texttt{@return}.

We refer to the sample Javadoc comments of Figure~\ref{fig:sampleprog} to exemplify and discuss possible results of this step, aiming to make the paper self contained.

The Javadoc comment \textit{@param limit an int, must be positive} (Figure~\ref{fig:sampleprog}, line~\ref{fig:sampleprog:tag:param}) specifies a precondition
on the value of parameter \texttt{limit}; 
any call of method \texttt{unwrapAndSave} that violates this precondition shall be considered invalid. \technique, via \jdoctor, renders this precondition contract as the formula \(limit > 0\).

From the Javadoc comment \textit{@throws NoMoreException if the drawer exceed the limit} (Figure~\ref{fig:sampleprog}, line~\ref{fig:sampleprog:tag:throws:nomore})  \technique extracts the postcondition contract

\noindent\(drawer.exceeds(limit) \implies retVal\ \text{instanceof}\ No\-More\-Exception.\)

\noindent According to this implication-style formula, if the antecedent \(drawer.exceeds(limit)\) is true, then the return value of the method must be an exception of type \texttt{NoMoreException}.

All postcondition contracts from \jdoctor are implication-style formulas. The antecedent of the implication is a condition on the values of input parameters of the method, and the consequent is an assertion on the return value of the method. 
The condition in the antecedent can also exploit predicates computed by calling methods of the program, if \jdoctor identifies those methods as suitable matches for the wording of the Javadoc comment. In this example, the condition in the antecedent, \(drawer.exceeds(limit)\), predicates on the input parameters \texttt{drawer} and \texttt{limit} based on method \texttt{drawer.exceeds}.

Hereon, for postcondition contracts, we refer to the antecedent and the consequent of the implication, as the \emph{guard} and the \emph{assertion} of the contract, respectively.

From the Javadoc comment \textit{@return false if the drawer already contains the gift, true otherwise} (Figure~\ref{fig:sampleprog}, line~\ref{fig:sampleprog:tag:return})  \technique extracts two postcondition contracts, 

\noindent\(drawer.contains(gift) \implies retVal=false,\) 

\noindent\(\neg drawer.contains(gift) \implies retVal=true,\) 

\noindent with guards \(drawer.contains(gift)\) and \(\neg draw\-er.con\-tains(gift)\), respectively, and assertions \(retValue=false\) and \(retVal=true\), respectively.

\subsection{The \technique Objective Functions}
\label{sec:objective:function}
\technique exploits the postcondition contracts
to generate test cases that address each of those contracts specifically. Namely, it aims to hit each postcondition contract with a dedicated test case, and prioritizes test cases that reveal possible violations of the contracts, that is, failures of the method under test. This means that, 
even if it cannot find any violation, it still aims to hit each of those contracts with a passing test case, i.e., a test case that satisfies the contract, as those test cases can assist regression testing tasks.

While pursuing the postcondition contracts, \technique exploits the precondition contracts as well, to ensure that it generates  valid test cases, which shall
 satisfy the preconditions.

\technique works in search-based fashion.  Thus, the core of the approach is to translate the postcondition contracts into the objective functions of an optimization problem over the space of the possible test cases for the class method under test. 
Given a postcondition contract $c$ defined for a method $m$, 
a solution of the optimization problem should be a test case that 
\begin{inparaenum}[(i)]
\item executes $m$ with parameter values that satisfy both the preconditions of $m$ and the guard of $c$, and
\item either violates (for a failing test case) or
satisfy (if no failing test case is found) the assertion of $c$.   
\end{inparaenum}
To this end, for a given postcondition contract $c$, \technique derives 
two separate objective functions, to which we refer as $\textsc{distance}_c(t)$ and $\overline{\textsc{distance}}_{c}(t)$, which capture to what extent a test case 
is distant from either satisfying or violating the contract $c$, respectively. 
Given a candidate test case $\hat{t}$, if $\textsc{distance}_c(\hat{t})$ (resp. $\overline{\textsc{distance}}_{c}(\hat{t})$) yields 0, it means that the test case $\hat{t}$ satisfies (resp. violates) the contract, and thus $\hat{t}$ is a solution that \technique is looking for.
In the case that \technique 
finds solutions for both objective functions, it  prioritizes the solution that violates the contract.

Below, we first define the objective function $\textsc{distance}_c(t)$, which addresses the satisfying test cases, and then define $\overline{\textsc{distance}}_{c}(t)$ by difference with the former one.

\paragraph{Searching for Contract-Satisfying Test Cases}
For a postcondition contract $c$, the objective function  $\textsc{distance}_c(t)$ associates any candidate test case $t$ with a value in the range [0, 1].
If $t$ satisfies the contract, then the objective function yields 0. Otherwise, it yields  positive values that grow up to~1 according to the extent to which the test is farther from satisfying the contract.
Thus, the objective function $\textsc{distance}_c(t)$ defines the optimization problem
\[\min_t\textsc{distance}_{c}(t)\]
of finding a test case on which the distance function evaluates to the minimum value 0. 

We now define $\textsc{distance}_{c}(t)$ in top down fashion, considering a  contract $c$ that refers to a method under test $m$.

First, we define $\textsc{distance}_{c}(t)$ as the composition of two sub-functions that separately address the subgoals of finding suitable inputs for the target method $m$, and  obtaining a suitable return value from $m$, i.e.
\[\textsc{distance}_{c}(t) = \frac{d^{in}_{c}(t) + d^{ret}_{c}(t)}{2}
\]
where:
\begin{itemize}
\item $d^{in}_{c}(t)$ yields 0 if the test case $t$ executes the method $m$ with inputs that satisfy both all preconditions of $m$ and the guard of $c$; otherwise, it yields positive values, up to maximum  1,
\item $d^{ret}_{c}(t)$ yields 0 if the test case $t$ executes the method $m$ such that the return value satisfies the assertion of $c$; otherwise,  it yields positive values, up to maximum 1,
\item both sub-goals provide  equal weight in the overall formula.
\end{itemize}

Next, we define $d^{in}_{c}(t)$ and $d^{ret}_{c}(t)$. To this end, 
\begin{itemize}
    \item let $\{pre_i\}$ be the set of precondition formulas that \technique extracted for the method $m$,
    
    \item let \(c\equiv\bigwedge_j guard_j\implies\bigwedge_kassert_k\) be the contract formula, where, with no loss of generality, we consider that both the guard and the assertion of the contract can be conjunctions of terms, $guard_j$ and  $assert_k$, respectively.   
\end{itemize}

We define $d^{in}_{c}(t)$ and $d^{ret}_{c}(t)$ as:

\[d^{in}_{c}(t) = \| \sum_{i} d^{in}_{pre\_i}(t) + \sum_{j} d^{in}_{guard\_j}(t)\|,\]
\[d^{ret}_{c}(t) =  \|\sum_{k} d^{ret}_{assert\_k}(t)\|,\]
where
\begin{itemize}
\item $d^{in}_{pre\_i}(t)$ evaluates a term $pre_i$ 
against the inputs with which the case $t$ executes the method $m$: it returns 0 if those inputs satisfy the term, or a positive value (maximum 1) otherwise, 
\item $d^{in}_{guard\_j}(t)$ addresses the terms $guard_j$ in similar way,
\item $d^{ret}_{assert\_k}(t)$ evaluates a term  $assert_k$ against the return value of method $m$: it returns 0 if the return value satisfies the term, or a positive value (maximum 1) otherwise, 
\item $\|.\|$ normalizes the results in the interval [0, 1]: $\|val\| = val/(1+val)$.  
\end{itemize}

In summary, $d^{in}_{pre\_i}(t)$, $d^{in}_{guard\_j}(t)$ and $d^{ret}_{assert\_k}(t)$ evaluate all relevant terms in the interval [0, 1], where 0 pinpoints test cases that suitably satisfy those terms. Then,
$d^{in}_{c}(t)$ and $d^{ret}_{c}(t)$ aggregate the results computed for the single terms,
to address contract $c$ as a whole.

We now finalize the definition by formalizing how we compute $d^{in}_{pre\_i}(t)$, $d^{in}_{guard\_j}(t)$ and $d^{ret}_{assert\_k}(t)$ for the possible types of terms. For a given $term$ (out of either $pre\_i$, $guard\_j$ or $assert\_k$), the computation consists in evaluating $term$ against
the inputs (resp. the return value) with which method $m$ is executed in the test case $t$. 

Any $term$ is a boolean predicate in the form $e_1 \bowtie e_2$, where
\begin{itemize}
\item $e_1$ and $e_2$ are arithmetic expressions over numeric literals and variables, and $\bowtie$ is some comparison operator; or

\item $e_1$ and $e_2$ are boolean expressions over boolean literals, variables and predicates, and $\bowtie$ is either $=$ or $\neq$; or

\item $e_1$ and $e_2$ are reference-typed values, including the possibility of \texttt{null} values, and $\bowtie$ is either $=$ or $\neq$.

\end{itemize}

For numeric terms the distance is
\[d_{(e\_1\ \bowtie\ e\_2)}(t) =
\begin{cases}
0, \quad\quad\quad\quad\quad\quad\quad\quad\quad \text{if}\ e_1(t) \bowtie e_2(t)\\
\||e_1(t) - e_2(t)| + \epsilon\|, \ \ \text{otherwise} 
\end{cases}\]
   
which yields $0$ if the comparison
$n_1 \bowtie n_2$ yields $true$. Otherwise, it is a positive value that
corresponds to the absolute value of the difference between $e_1$ and
$e_2$ according to the values passed to or returned from method $m$ in the test case $t$.
We add a small positive quantity $\epsilon$ to the difference between
$e_1$ and $e_2$, to handle the case of strict inequality comparisons.
For example, the precondition $limit > 0$ of method \texttt{unwrapAndSave} in Figure~\ref{fig:sampleprog}
is not satisfied by a test case $t$ that assigns parameter $limit$ to 0, and this is reflected by the
distance $d_{limit>0}^m(t)$ that yields
$\|0 - limit + \epsilon\| = \|0 - 0 + \epsilon\| = \epsilon / (1 + \epsilon)$, which is a non-zero albeit
small value.

This distance function favors test cases that better approximate the
solutions of the terms, aiming to yield a smooth objective
function, ultimately improving the search of an optimal solution. Referring again to the precondition $limit > 0$, a test case $t'$ in which $limit$ is $-1$
has a
distance $d_{limit>0}^m(t) = \|1 + \epsilon\|$ strictly greater than a
test case $t''$ in which $limit$ is $0$, and therefore the
search process will favor $t''$ over $t'$, thus progressing towards
the optimal solution.

For terms over boolean values, the distance follows the same formula, after turning the truth values of $e_1$ and $e_2$, to 1 or 0, if they are $true$ or $false$, respectively. Furthermore, we assign $\epsilon$ to 1, and dismiss the normalization operator. Thus the distance is 0 if the equality (resp. inequality) is satisfied, or 1, otherwise. For example, in the case of the postcondition \(drawer.contains(gift) \implies retVal=false,\) of method \texttt{unwrapAndSave}, 
both the guard and the assertion are terms over boolean values. In the case of the guard, the result of  $drawer.contains(gift)$ is implicitly compared with $true$: if the values of \texttt{drawer} and \texttt{gift} make
method \texttt{contains} return $true$, then the distance is 0; otherwise the distance is~1 because the guard is not satisfied.

Similarly, for terms over reference values, the distance is 0 if the equality (resp. inequality) is satisfied for the current reference values, or 1, otherwise. An example is the guard $drawer = null$ that \technique derives  from the Javadoc \emph{@throws NullPointerException if drawer is null} (Figure~\ref{fig:sampleprog}, line~\ref{fig:sampleprog:method:nullpointer}): if the value of parameter \texttt{drawer} is \texttt{null}, then then the distance is 0; otherwise the distance is 1 because the guard is not satisfied.

We remark that, even if the objective function of boolean-typed and reference-typed terms is flat (as it yields a binary outcome, either 0 or 1), the overall distance functions of many contracts is most often smoothed  
by the presence of  multiple terms that participate in the preconditions, guards and assertions of the contracts: The search engine receives guidance while incrementally satisfying all those terms.

\paragraph{Searching for Contract-Violating Test Cases}

For addressing test cases that violate the contract $c$, we define function $\overline{\textsc{distance}_c}(t)$ 
by adapting the definition given for contract-satisfying test cases.
The change is that we evaluate the return value of method $m$ according to the function
\[\overline{d^{ret}_c}(t) =  \prod_{k} d^{ret}_{\neg assert_k}(t),\]
which returns 0 if any of the terms $\neg assert_k$ is satisfied, meaning that some expectation in the contract assertion evaluates to $false$. Accordingly
 we define the optimization problem as
\[\min_t\overline{\textsc{distance}_c}(t) = \frac{d^{in}_c(t) + \overline{d^{ret}_c}(t)}{2},
\]
which looks for a test case that satisfies the preconditions of the target method and the guard of the contract, but violates any conjunctive term of the contract assertion.

\subsection{Generating Test Cases}
\label{sec:generating:test:cases}
\technique generates test cases by exploiting the objective functions of the target contracts with the search-based test generator \evosuite~\cite{Fraser:Evosuite:TSE:2013}.
To this end, \technique instantiates the objective functions 
in the form of \emph{evaluator programs}.
An evaluator program can be invoked at runtime with the current inputs and return value of a method under test, and yields a value in the range [0, 1] by computing the corresponding objective function.

Then, \technique leverages \evosuite
to build the test cases.
\evosuite generates test
cases by means of a genetic algorithm that randomly mutates and
combines a population of generated test cases to create new candidate solutions.
\technique makes \evosuite use
the evaluator programs (of the contract objective functions)
as the fitness functions of the genetic algorithm.
In this way, \evosuite will aim to progressively minimize the value returned by each evaluator
program (the fitness function), until it eventually succeeds in
generating a test case that makes the evaluator program return zero,
that is, a test case that satisfies the given contract.

\evosuite works by instrumenting the bytecode of the Java programs under test, such that it can track the execution of test cases in order to measure the fitness functions of interest. We adapted the \evosuite instrumentation such that, whenever a test case executes a class method that corresponds to a target contract, the execution will give control to the relevant evaluator program 
to trigger the computation of the fitness function for guards (before method execution) and assertions (after method execution).

\subsection{Contextualizing Test Cases on Contracts}
\label{sec:contextualizing:test:cases:contracts}
\technique further post-processes the test cases generated in the 
 test generation step, aiming to make them
\emph{contextualized on their corresponding focal contracts}.
In \technique, the \emph{focal contract} of a test case is the contract that corresponds to the objective function solved by the test case.
\technique generates a separate test case for  each solved objective function, and thus each test case corresponds to a single focal contract by design.

\technique post-process the test cases by 
\begin{inparaenum}[(i)]
\item adding the formulas of
the focal contracts as test oracles, and
\item tailoring the name of the test method
on the focal contract. \end{inparaenum}
It adds the contract assertion as \emph{assert oracle} after the invocation of the method under test.\footnote{Optionally, it may also add the preconditions and the contract guard as \emph{assume oracles}, before the invocation of the method under test. The assume oracles serve mostly as sanity checks, as the test case was generated such that those formulas are satisfied. Nonetheless,
they document the preconditions and the contract guard in the test code.} 
It synthesizes the test names by exploiting the natural language wording in the Javadoc of the focal contract:  it 
\begin{inparaenum}[(i)]
    \item uses the Stanford parser~\cite{manning2014stanford,de2006generating} to extract the relevant words of the description of both the guard and the assertion of the contract, that is, any name, pronoun, adjective, verb, adverb, preposition, coordinating and subordinating conjunction,   
    \item removes any character except letters and numbers,
    \item composes the name of the test method by concatenating in camel-case style: the string \texttt{test}, the name of the method under test, an underscore character, all relevant words of the assertion description, the string \texttt{If}, and all relevant words of the guard description.
\end{inparaenum}
In Figure~\ref{fig:jdttests} we already exemplified sample test cases generated by \technique.

\subsection{Limitations}
\technique depends on \jdoctor, which can sometimes misinterpret the natural language of the Javadoc comments,  and yield contract formulas
that imprecisely represent the documentation.
Also, \jdoctor can just fail to translate some Javadoc comments to contract formulas. 
Thus, even when \technique finds a test case that indeed violates a contract formula, the reported test failure could be a false alarm, if that formula happens to be affected by misinterpretations. The \jdoctor paper reports a precision of 92\%, but in our experiments (Section~\ref{sec:failures}) the impact on false-alarm reports is much higher. 
In the future, \technique could integrate with further, and possibly more precise, contract-synthesis technologies, in order to mitigate the annoying impacts of false alarm. 

Some contracts can be hard to satisfy. As any other search-based testing approach, \technique iterates the search process  until a maximum time
budget for the test search phase is reached.

\section{Experiments} \label{sec:experiments}
We empirically evaluated the effectiveness of  \technique in a set of experiments, based on a benchmark of 118 Java classes and 775 contracts extracted from the Javadoc in those classes. Thereby, we also compared \technique with a baseline test generator that targets code coverage in search based fashion. Below, we introduce the prototypes of both \technique and the baseline test generator, outline the research questions that drove the experiments, describe the benchmark, present the details of the experimental setting, and discuss our findings along with the possible threats to validity. We share \technique and our experimental results in a replication package.\footnote{https://anonymous.4open.science/r/Paper-AST2025-1342}

\subsection{Prototype and Baseline}
We implemented \technique in Java:
it extracts the contracts with \jdoctor, renders each contract as an objective function, generates test cases with a customized version of  \evosuite~1.2.1 (using the DynaMOSA heuristics) 
and post-processes the test cases to add names and oracles.

As \technique relies on \evosuite as back-end test generator, a crucial point in our experiments is to understand if it improves on the test cases that  
 \evosuite  can already yield with the classic code-coverage-based fitness functions.
Naively, we can 
simply augment those test cases by using the contracts from \jdoctor as oracles, following the test generation approach suggested in the \jdoctor paper (where they used the contracts 
as oracles in the test cases from Randoop~\cite{Blasi:Jdoctor:ISSTA:2018}).

We thus implemented a baseline test generator, to which we refer as \baseline. It generates test cases with \evosuite configured to use all classic fitness functions provided in the tool.\footnote{We configured \evosuite for using the fitness functions based on line coverage, branch coverage, exception coverage, mutation coverage, output coverage, and method coverage~\cite{Rojas:coverage:SSBSE:15,vogl2021evosuite}, with the DynaMOSA heuristics.
} It then adds the oracles by post-processing the test cases. For each call of a method $m$ and any postcondition contract $c$ of method $m$, it instruments an oracle as follows: before the call, it adds code that assigns a flag according to whether or not the parameters satisfy both the guard of $c$ and the preconditions of the $m$. After the call, if the flag is true, it crosschecks the assertion of $c$.

\subsection{Research Questions}
\label{sec:research:questions}

We aimed to answer the following research questions:
\begin{itemize}
    \item RQ1: Is \technique effective in testing the contracts extracted from the Javadoc of the software under test?
    \item RQ2: Does \technique detect actual failures?  
    \item RQ3: Is \technique original 
    in contextualizing test cases with focal contracts?
\end{itemize}

RQ1 aims to study the ability of \technique to generate test cases for contracts extracted from Javadoc comments, and detect possible violations of those contracts. We answer RQ1 by executing \technique against a benchmark of 775 contracts extracted from the Javadoc of 118 Java classes, and classifying which contracts \technique tests or does not test, along with the violations that it detects thereby. We compare the results of \technique with the ones of \baseline to quantify the extent to which \technique improves on the state of the art. For RQ1 we rely on the contracts that we could extract from Javadoc based on the technology of \jdoctor, without diving into whether or not those contracts might imprecisely represent some respective natural-language Javadoc specification. We consider those possible issues in RQ2.

RQ2 concentrates on the actual failures detected with
\technique. We manually inspected the failures revealed with the generated test cases, and classified which failures corresponded to actual violations of expected behaviors.
We submitted a set of failures to developers to collect  feedback. 

RQ3 investigates if the result of
crafting test cases that focus each on a single focal contract
is an original, distinctive characteristic of \technique,
or might extend to current approaches as well.   
Test cases that associate with single focal contracts can bring some benefits.
For instance, \technique exploits the knowledge of the focal contracts to define contextualized oracles and test names, which can improve the maintainability of the test suites~\cite{tufano2020unit,alagarsamy2023a3test,cem2003good}.
This can be harder if the test cases address multiple contracts.
Moreover, a test case that insists on multiple focal contracts may cause  failures of separate contracts to mask each other.  
We answer RQ3 by measuring the number of contracts tested in each test case from \baseline, and  comparing those numbers with the injective test-to-contract associations guaranteed with \technique. 

\subsection{Subjects}
We experimented with both \technique and \baseline against
the benchmark of 118  classes taken from 6 open-source Java systems.
The benchmark is the one available in the replication package of the seminal paper on the \jdoctor approach~\cite{Blasi:Jdoctor:ISSTA:2018,toradocu-ISSTA-software-artifact,toradocusrctestresourcesgoal} where they also made available the scripts to execute \jdoctor against those classes. 

Table \ref{tab:overallStats} lists the subject systems, the number of classes for each subject, and the number of precondition and postcondition contracts that \technique rendered (by calling \jdoctor) for the classes of each subject. It should be noted that we did not apply the filtering criteria discussed in the \jdoctor paper. We considered all contracts that \jdoctor yielded while analyzing the Javadoc of the classes. 
In total we considered 2,240 preconditions and 775 postcondition contracts. 
Column \textit{Id} assigns each subject with a corresponding identifier (in the range p1--p6) used for reference in the next tables below.

\begin{table}[t]
\centering
\caption{Statistics of the subjects\label{tab:overallStats}}
\begin{tabular}{clrrr}
\textbf{Id} & \textbf{Subjects} & \textbf{Classes} & \textbf{Preconds} & \textbf{Postconds} \\
\hline
p1 & Guava               &       19 &       203 &       72 \\
p2 & CommonsCollections4 &       20 &       348 &      211 \\
p3 & CommonsMath3        &       52 &       919 &      347 \\
p4 & GraphStream         &        7 &        73 &        7 \\
p5 & JGraphT             &       10 &        87 &        8 \\
p6 & Plume               &       10 &       610 &      130 \\\hline
&                        &      118 &      2240 &      775 \\
\end{tabular}
\end{table}

\subsection{Experimental Settings}

To control for the randomness of the search-based testing algorithm, we executed both \technique and \baseline 10 times against each class in the benchmark. 
Within both \technique and \baseline, we configured \evosuite version 1.2.1 with default options, e.g., population size set to 50.
We executed our experiments on a cloud resource set with Linux Ubuntu 18.04, 48 cores of Intel Xeon Gold 5120 and 169 gigabytes of RAM memory.

For the classes that corresponded to many postcondition contracts, 
we configured \technique to iterate
its back-end test generator
against batches of 10 contract fitness functions at a time. In this way, we allowed for some degree of parallelism, i.e., we let the back-end test generator  address multiple contracts at once, while mitigating the risk that too many fitness functions might interfere with the population size set for the genetic algorithm  of the test generator. We set the maximum time budget of each iteration of the back-end test generator, ranging from a minimum of 60 seconds, when executing with a single fitness function, and increasing according to the formula \(35*NumberOfFitnessFunctions\) seconds, when executing with batches of fitness functions of increasing size, up to a maximum of 350 seconds with 10 fitness functions.

For fairness, in the experiments in which we executed the baseline \baseline to pursue test generation in coverage-driven fashion, we set for each class a time budget equivalent to the sum of time budgets \technique used for that class. 
For example, for a class for which \technique should pursue 22 contract fitness functions, it would execute three iterations of the back-end test generator, with the fitness functions grouped across batches of size 10, 10 and 2, respectively, and
maximum time budgets set to 350, 350 and 70 seconds, respectively. For that class, 
we would execute \baseline with maximum time budget set to 770 seconds.

\subsection{Results on RQ1: Contract Testing}

We aimed to quantify the effectiveness of \technique in generating test cases that address contracts, and to what extent it improves on the \baseline baseline.
The RQ1 study focuses  on the ability of the test generation approach to provide test cases that  address the behaviors specified in the available contracts with suitable inputs and suitable oracles,
and not on the precision of extracting contracts from Javadoc.
Here, we relied on the postcondition contracts that \jdoctor could indeed extract (cfr.\ Table~\ref{tab:overallStats}) for the target classes.

For each postcondition contract available for each class of the considered benchmark, we measured whether or not a test generator (either \technique or \baseline) is able to test that contract, with  reasonable level of repeatability out of 10 executions. 
In detail, 
we first record whether the test generator \emph{hits} or \emph{misses} the postcondition contracts, 
for each execution of a test generator against a class. 
We record that a test generator hits a postcondition contract $c$ of method $m$, if the test generator yields a test case that executes the method $m$ with parameter values that satisfy both the guard of  contract $c$ and any precondition associated with the method. Otherwise, we record that the test generator misses the contract $c$.

Namely, in the case of \technique, we refer to the test cases that \technique explicitly generated with reference to the fitness function of the contract $c$, and we crosscheck (as a sanity check) that  those test cases execute method $m$ while satisfying both all preconditions of $m$ and the guard of $c$. Here we do not care of the assertion, as the contract $c$ is hit either if the test case does or does not reveal a failure. If \technique did not generate test cases while addressing the fitness function of $c$, we record that \technique misses the contract $c$. In the case of \baseline, we crosscheck the guard of $c$ and the preconditions of $m$ against each execution of method $m$ in any generated test case, and record that \baseline hits $c$ if we find that
the guard is satisfied at least once, or we record that \baseline misses $c$ otherwise. 

Next, we aggregate the data recorded in the 10 executions of each test generator. We report that a test generator tests the contract $c$, if it hits $c$ in at least 5 out of 10 executions. Otherwise, we conclude that, even if test generator hits the contract $c$ sometimes, its chances are low, and we conservatively infer that the test generator may rather miss $c$ most often.
Furthermore, we report that a test generator detects the violation of a contract $c$, if it yields a test case in which the contract oracle fails
in the majority of the executions in which it hits the contract.

\begin{table}[t]
    \caption{Tested contracts\label{tab:tested:contracts}}
    \centering
    \begin{tabular}{lp{45pt}| rrrrrr | r}
\bf Hit by & \bf  Outcome& \bf p1 & \bf p2 & \bf p3 & \bf p4 & \bf p5 & \bf p6 & \bf Total \\\hline\hline

\multirow{2}{18pt}{\technique} 
    & pass & 1 & 28 & 45 & 0 & 0 & 4 & 78  \\
    & alarm & 0 & 0 & 9 & 0 & 0 & 0 & 9  \\\hline
\multirow{4}{18pt}{Both} 
    & pass both & 48 & 80 & 195 & 4 & 6 & 68 & 401  \\
    & alarm both & 3 & 21 & 24 & 2 & 1 & 25 & 76  \\
    & alarm \technique & 7 & 5 & 5 & 1 & 1 & 3 & 22 \\
    & alarm E+J & 0 & 0 & 0 & 0 & 0 & 0 & 0 \\\hline
\multirow{2}{18pt}{E+J} 
    & pass & 1 & 0 & 1 & 0 & 0 & 0 & 2 \\
    & alarm & 0 & 0 & 0 & 0 & 0 & 0 & 0  \\\hline\hline
&  & 60 & 134 & 279 & 7 & 8 & 100 & 588  \\
    \end{tabular}
\end{table}

Table~\ref{tab:tested:contracts} reports the results
across the 6 analyzed projects. The first two rows below the table headers report how many contracts were tested by \technique, but not by \baseline, and for how many of those \technique reported test cases that passed (outcome  \textit{pass}) or detected a violation (outcome \textit{alarm}) in most cases, respectively.
The next four rows are for the contracts that both \technique and \baseline tested, and report: how many of those either passed or raised alarm with both techniques, how many raised alarm only with \technique, and no contracts (row \textit{Both:alarm E+J}) for which only  \baseline raised alarms.
The last two rows indicate that only 2 contracts were tested only by \baseline, none with violations.

The data indicate that \technique tested 76.6\%,  $(588 - 2) / 775$, of the postcondition contracts in the benchmark, stably across  10 repetitions of the experiments. In particular, it tested 87 contracts that code-coverage-driven test generation left untested.
Even more important, \technique uniquely revealed 31 contract violations (9 on uniquely hit contracts, and 22 on contracts hit by both techniques) in addition to the 76 failures that both techniques could reveal.

In summary, these results properly support our research hypothesis that code-coverage-oriented test generators may easily incur untested behaviors, and that 
generating test cases by explicitly addressing the contracts is effective in contrasting this phenomenon. 
Indeed \technique tested a significant portion (14.8\%,  $87/588$) of contracts for which \baseline failed to test the corresponding behaviors, and tested an additional share of 3.7\% ($22/588$)
 contract-violation behaviors that \baseline did not test. In total, \technique tested a portion of 18.5\% ($109/588$) of behaviors that \baseline did not test. 
These results also indicate that
addressing the contract violations explicitly (cfr. the distance function \(\overline{\textsc{distance}_c}(t)\) of \technique) is effective for detecting failures: whereas code-coverage-oriented test generators detect failures just by their coincidental relations with executing code elements, \technique can successfully discriminate between executions that traverse the same code with different outcomes.

Finally, the 2 contracts that \baseline uniquely tests indicate the possible complementarity of the two approaches. Our inspection of those cases revealed that, as  expected, some relevant inputs can correspond to flat regions of the fitness landscape pursued with \technique, but can be captured with \baseline as they increase code coverage. In future work, we aim to explore the possibility of combining the two approaches to exploit their mutual  benefits. 

\subsection{Results on RQ2: Detection of Actual Failures}
\label{sec:failures}
We manually inspected a failing test case for each
of the 107 contracts for which \technique detected a violation (cfr.\ Table~\ref{tab:tested:contracts}, $9+76+22$ contracts).
We classified those violations as either \emph{actual} or \emph{false  alarms}, as we report in Table~\ref{tab:alarms}.

We marked a false alarm if the manual inspection revealed that, regardless of the failure of the test case generated based on derived contract formulas, the behavior of the program under test  complies with the natural-language counterpart in the Javadoc. This may happen because \jdoctor can sometimes misinterpret the natural language descriptions~\cite{Blasi:Jdoctor:ISSTA:2018}.\footnote{ 
A peculiar set of false alarms happen because some test cases assign floating-point inputs with $NaN$ values. $NaN$ stands for not-a-number, which in Java can  be the result of invalid operations like division-by-zero. The $NaN$ values obey to arithmetic rules that may mislead some contract oracles. For example, if a test case assigns the input $a$ to $NaN$, an oracle that predicates $retValue=a$ would fail against a program that correctly returns the value of variable $a$, as the oracle check $NaN=NaN$ returns $false$.
}

\begin{table}[t]
    \caption{Analysis of the contract-violation alarms\label{tab:alarms}}
    \centering
    \begin{tabular}{l|rrrrrr|r}
             & \bf p1 & \bf p2 & \bf p3 & \bf p4 & \bf p5 & \bf p6 & \bf Total    \\\hline
    \bf Actual alarms  &  8 & 18 &  5 &  1 &  2 & 11 &  45  \\
    \bf False alarms     &  2 &  8 & 33 &  2 &  0 & 17 &  62  \\
    \end{tabular}
\end{table}

\begin{table}[t]
    \caption{Actual failures report classification\label{tab:failures:report}}
    \centering
    \begin{tabular}{p{55pt}| rr }
 Fix status & Guava & CommonsCollections4  \\\hline
    Fixed newly      & 0 &  2  \\
    Fixed in the past & 2 &  1 \\
    Will fix      & 0 &  4 \\
    Waiting reply & 0 & 11\\
    Won't fix     & 6 &  0 \\\hline
             & 8 & 18 \\
    \end{tabular}
\end{table}

The results in Table~\ref{tab:alarms} show
that 42\% ($45/107$) of the identified contract violations were indeed  actual alarms, i.e, those test cases reveal actual failures of the software. Worth noting, only two of those failures were reported by Blasi et al.~\cite{Blasi:Jdoctor:ISSTA:2018} in their experiments with \jdoctor and \textsc{Randoop} against the same benchmark of Java classes.

We aimed to collect feedback on those failures by reporting them to the developers of the subject systems. We could do that only for the subjects Guava (p1) and Commons\-Collections4 (p2), because, at the time of writing, the projects of the other four subjects, either have been discontinued, or 
moved to new  versions where the classes of our benchmark do not exist anymore.
Table~\ref{tab:failures:report} summarizes the status of the developer feedback that we collected so far on our 26 failure reports related to  Guava and Commons\-Collections4. We include all links to the issue reports in the replication package.\footnote{\url{ https://anonymous.4open.science/r/Paper-AST2025-1342}}

Indeed 9 out of 26 failures were confirmed: 
We found 2 failures of Guava and 1 of Commons\-Collections4 that had already been fixed by patches provided in the past, while the developers of Commons\-Collections4 promptly provided the fixes for 2 unknown failures, and already confirmed upcoming fixes on other 4 unknown failures. 
For 6 failures of Guava, developers do not plan fixes, as those failures map to lack of documentation that is considered fine anyway.
On 11 failures of Commons\-Collections4  we did not receive any feedback yet.

\subsection{Results on RQ3: Test-to-Focal-Contract Associations}

We counted the number of focal contracts for each generated test case. The test cases from  \technique hit each a single contract by construction.
The test cases from \baseline, aiming to contribute in maximizing code coverage, often end up with testing multiple contracts  in a test case, or even none of the target contracts in many test cases. 

For \baseline, we first counted how many test cases generated in the 10 repetitions of our experiments
do not test any contract. We found that this happens for 67\% of  58,366  test cases, in many cases because the provided inputs miss the preconditions. For the remaining 33\% of the test cases, we found that 
34\% of those test cases hit between 2 and 15 contracts each. Thus they have no specific focal contract.

We conclude that only \technique has the distinctive merit of
contextualizing the test cases each on its own specific focal contract. This brings the benefits on avoiding masking between failures that occur concomitantly 
and attributing meaningful names to the test cases, as we already discussed above.

\subsection{Threats to Validity}
The random choices that underlie the genetic algorithm of the back-end test generator \evosuite threaten the internal validity of our conclusions, as in principle every run of the test generators might lead to test or not to test different sets of contracts.
We mitigated this threat by repeating all experiments 10 times, and interpreting the findings only for the contracts that could be steadily tested in at least half of the replicas. 

Another threat relates with the correctness of our implementation of \technique.
We set the \technique  test cases  with sanity checks that 
the test outcome is different than  the satisfied objective function. We did not observe any warning.

External validity is threatened  as our results may not generalize beyond the specific sample of subjects considered in our experiments, e.g., for other Java systems,  for Javadoc comments written by different teams, or programs encoded with programming languages different than Java or documentation systems other than Javadoc. 
We believe that our results are promising,
 but we are also aware that our findings shall be further crosschecked with replication studies in the future.

\section{Related Work}
\label{sec:related}
The contracts that we addressed with \technique 
could be  
instrumented within the program under test as
checking code  for the preconditions and the postconditions from \jdoctor, 
at the beginning and the end of the corresponding methods' body, respectively. 
Then we could apply formal verification tools or existing test generators~\cite{cordeiro2018jbmc,Tillmann:pex:TAP:2008,Fraser:Evosuite:TSE:2013}.

Such approach leads to
naive solutions with many drawbacks. For instance, we tried to use \jbmc, which realizes bounded model checking for Java programs~\cite{cordeiro2018jbmc}. For the method \texttt{unwrapAndSave} in Figure~\ref{fig:sampleprog}, \jbmc pinpointed the 2 bugs that we discussed, but also 6 additional false alarms.
The false alarms arise because bounded model checking assumes unconstrained inputs, e.g., \jbmc assumed that callers may pass a \texttt{drawer} object in which the underlying data structure includes null pointers; but such an object cannot be instantiated by using the actual methods of class \texttt{Drawer}. 

Similar issues occur for test generators based on directed testing~\cite{korel1996assertion,sen2006cute,Tillmann:pex:TAP:2008,lyu2020automated,caballero2015checking}.

For instance, the tool Pex~\cite{Tillmann:pex:TAP:2008} addresses the assertions with dynamic symbolic execution. Upon detecting path conditions that violate the assertions, it instantiates test cases that demonstrate the corresponding failures, 
by assuming the possibility of freely manipulating the internal fields of the input objects. It thus
results in the same false alarms as \jbmc.
The false positive rates of \jbmc and \pex (75\% in the example) are  high for most  programs, making those approaches hardly viable as solutions to our problem. 

We can also implement directed testing with \evosuite, executing it directly against the programs instrumented with the contracts. \evosuite accounts for the program APIs. This  way of  using \evosuite incurs hybrid guidance from both the contracts and the  program branches: it
yields (unfocused) test cases that may satisfy each several or no contracts.

We can of course imagine several ways of tweaking or combining the above approaches to better suit our problem. Nonetheless, we remark that the possibility of attempting other solutions, rather than \technique, 
does not per-se  affect the novelty of our idea.
We maintain that our paper is unprecedented in proposing to generate test cases that \emph{constructively address the functional behaviors documented in code comments}, and  we provided novel insights by evaluating this idea in comparison with classic test generations based on code coverage.
We leave  for future work the comparison between \technique and other possible ways of designing a similar solution, including the possibility of using other metaheuristic algorithms~\cite{Tracey:annealing:ISSTA:1998}.

\technique has also relations with the work on Athena and A3Test~\cite{tufano2020unit,alagarsamy2023a3test} that also aim at generating unit test cases with meaningful names and assertions.
Those approaches apply deep learning with training on test cases written by developers, and thus generate oracles by similarity with other test cases.  
Other approaches derive oracles by observing program executions, via either  differential testing~\cite{davis1981pseudo,feldt1998generating,mcminn2009search}, or metamorphic relations between different executions~\cite{murphy2009automatic,murphy2009using,guderlei2007statistical}, or inferring invariants with dynamic analysis~\cite{Wei:BetterContracts:ICSE:2011,Wei:2011:STF:2190078.2190095}. 
\technique differs from all those approaches in that it addresses the expectations documented in the programs under test.

\section{Conclusions}
\label{sec:conclusions}
Augmenting programs with Javadoc documentation is  common practice for Java programmers. This paper presented a novel approach, \technique, to generating test cases that constructively address the method contracts documented in Javadoc.

\technique instantiates search-based testing with objective functions that represent the target contracts, aiming to generate test cases that either execute the program behaviors documented as contracts, or reveal violations of those contracts. It yields test cases with names and oracles tailored on the target contracts.
Our experiments provide empirical evidence
that \technique successfully 
tests a significant portion of relevant program behaviors that classic code-coverage-driven test generators may often miss, and effectively reveals  failures.

In the future we aim 
to experiment \technique with other Java systems and Javadoc documentation written by different teams, and we aim to implement \technique for programming environments different from  Java and
Javadoc.
\bibliographystyle{IEEEtran}
\bibliography{otherbib,bibliography}

\end{document}